# Enhancing Thermoelectric Properties of Isotope Graphene Nanoribbons via Machine Learning Guided Manipulation of Disordered Antidots and Interfaces

Xiang Huang[1], Shengluo Ma[1], Haidong Wang[2], Shangchao Lin[1], C. Y. Zhao[1], Hong Wang[3], and Shenghong Ju[1, 3, *]

[1] China-UK Low Carbon College, Shanghai Jiao Tong University, Shanghai, 201306, China

[2] Key Laboratory of Thermal Science and Power Engineering of Ministry of Education, Tsinghua University, Beijing, 100084, China

[3] Materials Genome Initiative Center, School of Material Science and Engineering, Shanghai Jiao Tong University, Shanghai 200240, China

**Abstract**

Structural manipulation at the nanoscale breaks the intrinsic correlations among different energy carrier transport properties, achieving high thermoelectric performance. However, the coupled multifunctional (phonon and electron) transport in the design of nanomaterials makes the optimization of thermoelectric properties challenging. Machine learning brings convenience to the design of nanostructures with large degree of freedom. Herein, we conducted comprehensive thermoelectric optimization of isotopic armchair graphene nanoribbons (AGNRs) with antidots and interfaces by combining Green's function approach with machine learning algorithms. The optimal AGNR with *ZT* of 0.894 by manipulating antidots was obtained at the interfaces of the aperiodic isotope superlattices, which is 5.69 times larger than that of the pristine structure. The proposed optimal structure via machine learning provides physical insights that the carbon-13 atoms tend to form a continuous interface barrier perpendicular to the carrier transport direction to suppress the propagation of phonons through isotope AGNRs. The antidot effect is more effective than isotope substitution in improving the thermoelectric properties of AGNRs. The proposed approach coupling energy carrier transport property analysis with machine learning algorithms offers highly efficient guidance on enhancing the thermoelectric properties of low-dimensional nanomaterials, as well as to explore and gain non-intuitive physical insights.

* Corresponding author: shenghong.ju@sjtu.edu.cn (S.H. Ju)

## 1. Introduction

Thermoelectric materials have drawn tremendous attention as they can directly convert waste heat to electricity [1-4]. The efficiency of thermoelectric material is characterized by dimensionless figure of merit *ZT*, i.e. $ZT = \sigma S^2 T/(k_p + k_e) = Z_p T/k$, where $\sigma$ is the electrical conductivity, *S* is the Seebeck coefficient, *T* is the absolute temperature, $Z_p$ is the thermoelectric power factor, $k_p$, $k_e$ and $k$ are the lattice, electronic and total thermal conductivity, respectively. The optimization of *ZT* for thermoelectric materials involves multi-parameter synergies. Ideal thermoelectric materials need to have high electrical conductivity and Seebeck coefficient, but low thermal conductivity [5]. However, the fact is that these properties are deeply coupled with each other, making it difficult to optimize the thermoelectric properties of materials.

Advanced materials/structures such as SnSe [6], clusters [7, 8] and graphene [9] have made promising progress in terms of thermoelectric properties over the past few years. Some bulk materials including SnSe and clathrates lead to strong crystal orientation or structure anisotropy and have large *ZT* in certain directions. However, the high *ZT* of these materials is strongly dependent on the temperature [10-12]. For example, a recently reported hole-doped SnSe polycrystalline sample exhibited a *ZT* of roughly 3.1 at 783 *K*, but the *ZT* at 300 *K* is less than 0.5 [13]. Designing materials/structures with excellent thermoelectric properties at room temperature have broader applications, such as providing energy for wearable devices [14]. As the material size scale decreases from high dimensional to low dimensional, it will show some extraordinary properties [15]. Graphene has been considered as an ideal flexible low-dimensional thermoelectric material with the advantages of lightweight, non-toxicity, high mechanical flexibility, good thermodynamic stability, large electron mobility and scalable synthesis in manufacturing [9, 16, 17]. Pristine graphene hardly exhibits good thermoelectric properties due to its high thermal conductivity [18] and small electronic band gap [19]. However, one can reduce the thermal conductivity and open up the band-gap to increase the Seebeck coefficient by properly introducing the defects or topological design [20, 21]. Graphene with disordered edges [22], rough surfaces [23], interfaces [24], isotopes [25-27] and nanopores [28, 29] can enhance phonon scattering with the defects thus reducing lattice thermal conductivity. Furthermore, topological designs such as controlling the carrier transport direction width to form graphene nanoribbons (GNRs) [30] and inserting antidots [31] enable to increase the electronic bandgap and obtain large Seebeck coefficients. Among these approaches, isotope [32, 33] and antidot engineering [34, 35] are two common solutions for both theoretical and experimental implementation. The graphene isotope superlattices can be synthesized via chemical vapor deposition technology by pulsed signal controlled gases with different isotope carbon sources into the reaction chamber, which has the advantages of easy preparation and high reliability [36, 37]. Moreover, the scale regulation of GNRs and the preparation of nanopores/antidots are enabled by controllable ion beam etching [30, 34, 35, 38]. What's more, the thermal conductivity, electrical conductivity and Seebeck coefficient of GNRs are measurable by an eight-terminal device, details can be found elsewhere [39].

Although the design of GNRs thermoelectric devices has accumulated a wealth of theoretical and experimental methods, its properties optimization still faces challenges due to the variety of regulatory means and structural degrees of freedom (DOF). The traditional trial-and-error method has disadvantages such as high cost, time consumption and the probability of success is also uncertain [40]. Material informatics integrated with informatics algorithms for material structure optimization has been demonstrated superior to traditional empirical trial-and-error methods in the design of multi-degree-of-freedom thermal functional materials [41, 42]. It has high efficiency in thermal transport design [43-45], thermoelectric optimization [46-48] and thermal radiation design [49-51]. Moreover, some optimal structures or devices such as aperiodic GaAs/AlAs superlattice structure with low coherent phonon heat conduction [52] and highly wavelength-selective, multilayer nanocomposite selective thermophotovoltaic emitter [53] have been experimentally fabricated, which demonstrates the applicability and efficiency of the informatics algorithms. In this work, we proposed an approach coupling energy carrier transport analysis with machine learning algorithms to optimize the thermoelectric properties of isotope-substituted armchair graphene nanoribbons (AGNRs). Before optimization, the thermoelectric properties of AGNRs with different widths were investigated as they exhibit metallic or semiconductor characteristics. Then, we demonstrated the effectiveness of the approach for isotopically labeled AGNRs and analyzed the isotopic effects on thermal transport from the viewpoint of phonon vibrational modes. To further obtain the optimal structure with larger *ZT*, we designed the AGNR isotope superlattices with antidots at the interfaces, and discussed the contribution of antidot effects and isotope manipulations. Furthermore, we analyzed and evaluated the optimization ability and efficiency of machine learning algorithms including Bayesian optimization (BO) and genetic algorithm (GA).

## 2. Computational details and machine learning algorithms

*2.1 Computational details*

We constructed two types of AGNRs for thermoelectric performance optimization, as shown in Fig. 1. In a periodically isotope substituted AGNR (model A, see Fig. 1a), the size was controlled by length $M$ and width $N$, where $M$ was the number of the smallest repeating structural unit ($L$ was three times the length of the C-C bond with 0.426 nm) in the length direction, and $N$ was the number of carbon atoms in the width direction. Parameter $M$ was uniformly set as 60 ($L$ was 25.56 nm) to maintain a proper system size for calculation. As the parameter $N$ equals $3p$, $3p+1$, or $3p+2$ (where $p$ is an integer), AGNRs can be classified into three families showing different electronic properties [54, 55]. We referred to the AGNR with $N$ carbon atoms width as the $N$-AGNR. To conduct more realistic calculations, the edges of all the AGNRs in the study were passivated by hydrogen atoms [56, 57]. For different candidate structures in model A, the binary flags "1" and "0" were adopted as the descriptors for C-12 and C-13, respectively [43, 58, 59]. Furthermore, we considered the aperiodic AGNRs with isotopes and antidots (model B, see Fig. 1b) which was the more realizable structure in experiments. The minimum cell was composed of 4×13 lattice units cells of GNRs, and the antidots were directly set at the isotope interfaces. Similarly, binary flags were used as the descriptors. For example, the unit with C-12 was encoded as "1", the adjacent unit with C-13 was encoded as "0", and the antidot was located at the interface between the two units.

Before the calculation of thermoelectric properties, all constructed AGNRs were optimized using Tersoff potential until the maximum interatomic force became less than 0.001 meV/Å [60]. Then, the electron and phonon transmission functions were obtained by using the Nonequilibrium Green's function (NEGF) with Landauer formalism (See details in section S1 of supplemental materials) [61-66]. As shown in Fig. 1a, the scattering region was divided into three parts: central region, left and right leads. The left/right lead was the extension of the left/right electrode, to screen out the perturbations from the scatter inside the device [67]. In consequence, the left and right leads remained pristine C-12, and the central region was the designing area. For electron transport properties, the self-consistent charge density functional tight binding scheme (SCC-DFTB) was used to express the interaction among $1s$ orbital of H and $2s$ and $2p_{x/y/z}$ orbitals of C [68-70]. SCC-DFTB is an approximate DFT approach that provides ab initio accuracy for systems for which it is parameterized [71]. Moreover, the chemical potential energy $\mu$ was set from -1.5 eV to 1.5 eV, which was large enough to obtain the maximum *ZT* [58]. As for the calculation of phonon transmission function, optimized Tersoff potential was applied to C–C, C-H and H-H interatomic interactions [60]. The optimized potential parameter sets reproduce the graphene phonon dispersions more accurately than the original parameter sets, which has been widely applied in the investigations of graphene isotope labeling [72], topological [73] and interfacial [74, 75] phonon engineering. All calculations were performed at 300 *K*. Since the length of GNR (*L*<30 nm) in our system is much smaller than the electron mean free path, the interaction between electron and phonon was ignored [76].

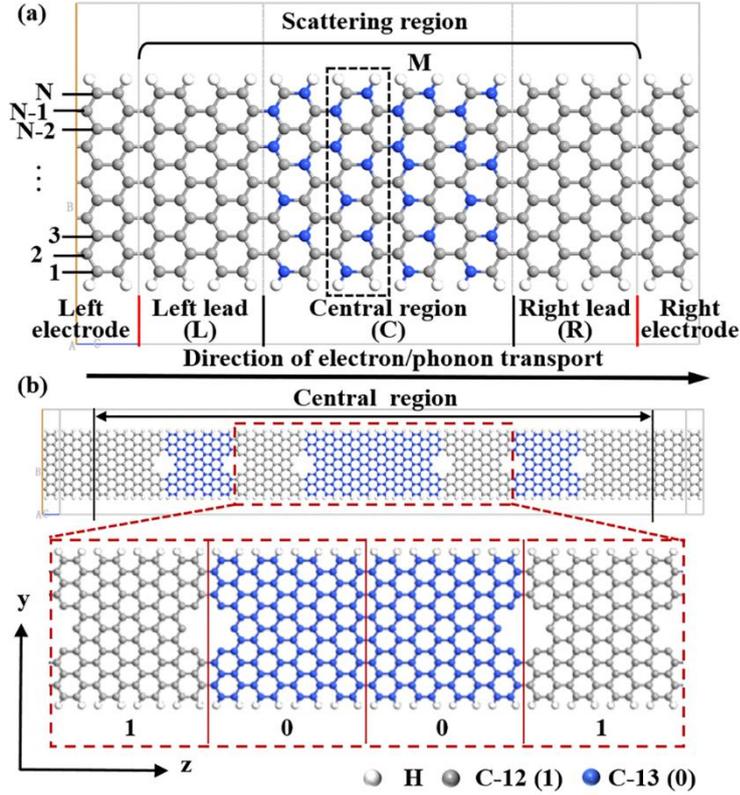

**Fig. 1.** The schematic diagram of the AGNRs. (a) Periodically isotope substituted AGNR. (b) Aperiodic AGNR with isotopes and antidots. The antidots were set at the isotope interfaces.

*2.2 Machine learning algorithms*

Optimization algorithms of BO [77] and GA [78] were used to improve the thermoelectric properties of AGNRs respectively for comparing their optimization ability and efficiency. BO uses Bayesian theory and Gaussian process to find the optimal AGNR [79], and the process can be described as:1) training the Bayesian model by several initial calculated structures; 2) selecting promising candidate structures according to the training model to calculate accurate *ZT*; 3) adding new calculated structures and properties to the dataset for next loop; 4) repeating the above process until the number of iterations is reached. The Bayesian linear regression model was used for the prediction of *ZT*,

$$ZT = w^T \varphi(x) + \varepsilon, \tag{1}$$

where *x* is a *d*-dimensional vector (*d* equal to the number of descriptors), *w* is a *D*-dimensional weight vector, $\varphi$ is the feature map including *D* basis functions, and $\varepsilon$ is the noise subject to Gaussian distribution with the mean of zero and the covariances $\sigma^2$.

GA is a heuristic search algorithm inspired by the biological evolutionary process, which initially starts with a broad set of solutions that are represented by chromosomes called populations [80]. The solutions obtained from a single population are retained and used to reproduce and form new populations. Motivated by optimism, the genetic algorithm expects to retrieve a new generation that is better and superior to the previous one. The main process based on GA includes four steps[81]: 1) Initialization, generating random populations of several chromosomes and evaluating them by using the NEGF method to calculate *ZT* as the fitness index; 2) Selection, selecting two paternal chromosomes from the current populations based on fitness evaluation; 3) Crossover, exchanging some genes in two chromosomes to construct new chromosomes for the next generation; 4) Mutation, generating new offspring due to a certain probability of mutation of some genes in the chromosome and placing the new offspring into the newly generated population. Here, the optimization with "0" and "1" as discrete codes was carried out, the initial population size was 400, and the variation probability was set as 0.012.

Figure 2 shows the schematic of improving the thermoelectric properties of AGNRs via the method combining NEGF and machine learning. We utilized the Quantum ATK software [67] for the analysis of electron/phonon transport of AGNRs, and employed the open-source libraries COMBO [77] and Scikit-opt [78] for the algorithms of BO and GA, respectively.

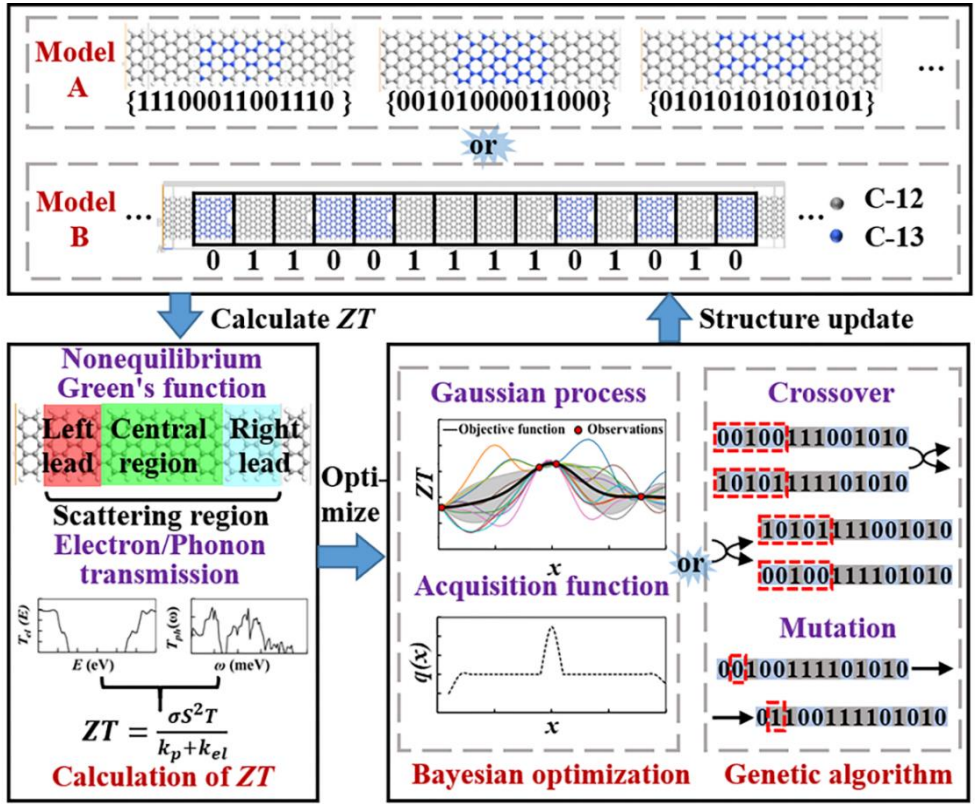

**Fig. 2.** Schematic of improving the thermoelectric properties of AGNRs via the method combining NEGF and machine learning.

## 3. Results and discussion

Three types of AGNRs (7-AGNR, 8-AGNR and 9-AGNR) were established first to explore the effect of AGNR width on thermoelectric properties (see details in Section S2 of supplemental materials). The results of phonon transport properties show that the thermal conductivity of AGNR increases monotonically in the width range of $N$ from 7 to 9 at 300 K (Fig. S1). In the case of narrower AGNRs, the edge-localized phonon modes play a dominant role, which is weakened with the increase of AGNR width and leads to increased thermal conductivity [82]. For electron transport properties (Fig. S2), the 8-AGNR with $N=3p+2$ exhibits metallic properties, which has the smallest energy bandgap and peak of Seebeck coefficient [83]. However, 7-AGNR and 9-AGNR behave as semiconductors, the energy bandgap and peak of Seebeck coefficient are larger than that of metallic graphene respectively and are inversely proportional to the width [84, 85]. The thermoelectric properties of AGNRs (Fig. S3) are affected by the coupling of electrical conductivity $\sigma$, Seebeck coefficient $S$ and thermal conductivity $k$. It is worth mentioning that the power factor $Z_p$ is sensitive to the dependence of the electrical conductivity, whose peak positions correspond exactly to the electron conductance transitions from one step to another, rather than the positions of the $S$ peak [66]. This observation is consistent with the Cutler-Mott theory, i.e. $Z_p \propto d[ln\,\sigma(\mu)]/d[\mu]$ [86]. The $ZT$ peak of 7-AGNR, 8-AGNR and 9-AGNR are 0.245, 0.242 and 0.147, respectively. In addition, ignoring lattice thermal conductivity, the $ZT_e$ peak of 7-AGNR and 8-AGNR are about 706.264 and 21.204, which are 2882.6 times and 87.6 times of the corresponding $ZT$, respectively. The difference between $ZT_e$ and $ZT$ reflects the significant contribution of lattice thermal conductivity to the total thermal conductivity. Besides, the AGNR with $N=3p+1$ is more sensitive in the same sequence (the integer $p$ is equal). Based on the above results, we chose the AGNRs with $N=3p+1$ for the following discussions.

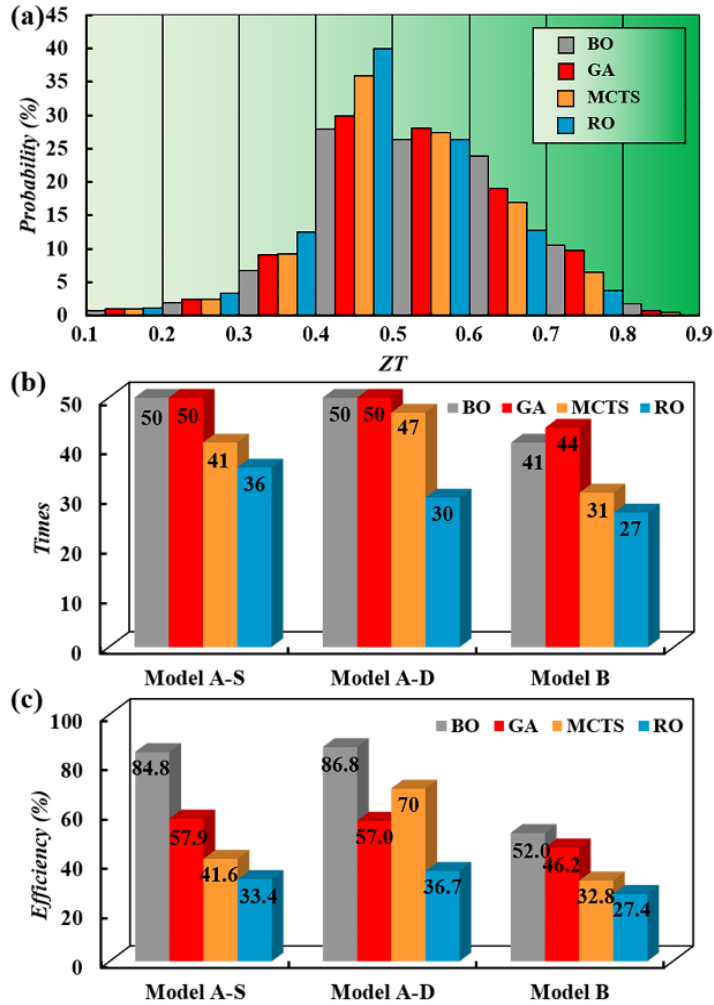

**Fig. 3.** Comparison of the optimization ability and efficiency of different algorithms in 50 rounds of optimization with different initial choices of AGNR structures. (a) $ZT$ values of optimized AGNR structures (model B) in the worst round. (b) and (c) Optimization ability and optimization efficiency of each algorithm.

To evaluate the optimization ability and efficiency of different algorithms, 50 rounds of optimization were performed with different initial choices of AGNR structures, and 2000 structures were calculated in each round. The optimization ability was evaluated by the number of times to find out the AGNR structures corresponding to the top-5 $ZT$ values, and the optimization efficiency $\eta$ is defined as $\eta=(2000-i)/2000$, where $i$ is the first time to find the top-5 AGNR structures in $ZT$ after calculating $i$-th structures in a round. The larger the $\eta$, the higher the optimization efficiency of the algorithm. Here, we additionally performed Monte Carlo tree search (MCTS) and random optimization (RO) algorithms for comparison (See details in Section S3 of supplemental materials). The convergence of the curves in Fig. S4 visually illustrates the optimization ability of each algorithm. It indicates that the BO and GA have stronger optimization ability than other algorithms. However, for the same algorithm, the optimization ability decreases in model B. To gain insights into the performance of each algorithm, we further counted the distribution of $ZT$ values of optimized AGNR structures (model B) in the worst round, as shown in Fig 3a. Most of the $ZT$ values are concentrated in the middle region from 0.4 to 0.6. However, the AGNR structures suggested by BO are in the high $ZT$ region (0.6~0.9), and the advantage is more obvious with $ZT>0.8$. As for the other three algorithms, the optimization ability ranking according to the number of suggested high $ZT$ structures is GA, MCTS and RO. Figure 3b shows the optimization ability of each algorithm in different systems. We performed the optimization of periodic AGNR structures within single and double units in model A, respectively, named model A-S and model A-D. With 50 rounds of optimization, both BO and GA can find out the top-5 structures of $ZT$ in model A, but GA behaves a little better in model B even if not find out in every round. The optimization efficiency of each algorithm is depicted in Fig. 3c, and it shows that BO has a higher optimization efficiency which means that it is possible to calculate fewer structures to discover the optimal structure with higher thermoelectric performance. To summarize the optimization efficiency here we can conclude that, with the benefit from Bayesian theory and Gaussian processes, the BO has high optimization efficiency and requires fewer iterations to obtain a better result. Although the selection, crossover, and mutation processes in GA are randomized, their purpose is to improve fitness. By continuously combining the information from the previous iteration, it gives the random process the direction that controls the convergence of the algorithm. Therefore, GA has a stronger optimization ability. MCTS is a random best-first tree search algorithm, which seeks an optimal balance between asymptotic convergence in a definite direction and randomness. Compare to RO, MCTS behaves better, even if not as well as BO and GA.

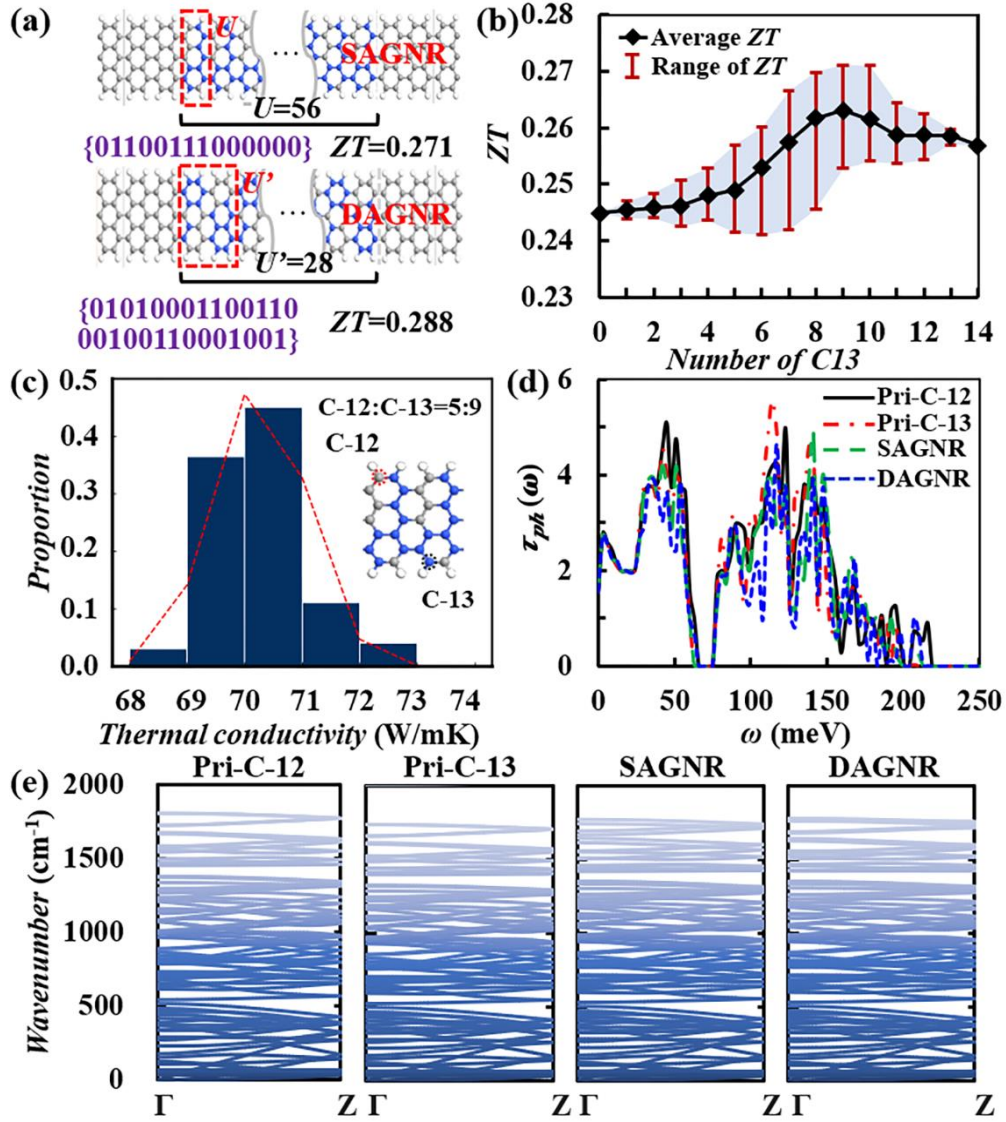

**Fig. 4.** Optimization of periodically isotope-substituted AGNR structures. (a) Optimal structures obtained from different regulating DOF. (b) The *ZT* of all candidate structures varies with the number of C-13, from the system of single unit regulating DOF. (c) Thermal conductivity distribution of AGNR candidates at a fixed C-13 concentration of 9/14. The red dashed line for fitting the normal distribution. (d) and (e) Phonon transmission function and dispersion relationship of different AGNR structures. The structures labeled Pri-C-12 and Pri-C-13 indicate that the pristine structures with all carbon atoms are C-12 and C-13 in their central region, the structures labeled SAGNR and DAGNR are the optimal structures from the systems of single and double units regulating DOF, respectively.

The effect of isotopes has a significant impact on phonon transport properties, while the effect on electron transport is almost negligible [25]. To investigate the influence of DOF of structural regulation on thermoelectric performance, we selected single and double units as minimum period cells, respectively, while the length of the 7-AGNR was kept constant. When varying quantitative ratios of C-12 and C-13 in periodic AGNRs, the obtained optimal structures are shown in Fig. 4a. The *ZT* of optimal structure from the system of double units regulating DOF (DAGNR) is 0.288, which is larger than that of SANGR (single unit regulating DOF). It indicates that when increasing the design degree of freedom, the AGNR with larger *ZT* can be expected to be found by rational design of optimization strategy. In Fig. 4b, we calculated all candidate structures ($2^{14} = 16384$) from the system of single unit regulating DOF to investigate the effect of C-13 concentration on *ZT*. Despite the consistent trend of average *ZT* ($Z_{ave}T$) and maximum *ZT* ($Z_{max}T$) with changing C-13 concentration, the distribution of isotopic carbon atoms is also important in tuning the thermal conductivity/thermoelectric properties of AGNRs. Figure 4c shows the thermal conductivity distribution of AGNR candidates at a fixed C-13 concentration of 9/14. The optimal structure (Type-II in Fig. S5a) has the smallest thermal conductivity of 68.11 W/mK and the maximum *ZT* of 0.271. However, the minimum *ZT* is 0.253 with the same C-13 concentration (Type-III in Fig. S5a), which is smaller than that of optimal structures with most other C-13 contents. To clarify this concern, the additional arbitrary C-13 concentration AGNR structure of {00110111011001} with the largest thermal conductivity was adopted for the analysis (Type-I in Fig. S5a). For the phonon transport calculations, the electrodes and extension regions are uniformly set to C-12. The Type-I AGNR intuitively provides a continuum path of C-12 for phonons to coherently propagate, thus having the largest thermal conductivity. Similarly, although the Type-III AGNR fixes the C-13 concentration, it still maintains a continuous path of C-12 for coherent phonon propagation and has the maximum thermal conductivity at this concentration. The Type-II AGNR is the optimal structure, whose C-13 atoms form a continuum interface barrier in the direction perpendicular to the transport, suppressing the propagation of phonons. The above results are consistent with previous studies on the thermal conductivity of Si/Ge interface structures [43]. Additionally, phonon transmission and dispersion relations of pristine C-12, pristine C-13, SAGNR and DAGNR structures were calculated separately for the analysis of the role of isotope atoms, as shown in Fig. 4d and 4e. Compare to pristine structures, the reasonable C-13 concentration and distribution in SAGNR and DAGNR structures form a continuum interface barrier and suppress the propagation of phonons with the reduced group velocity, which is appearing in a wide frequency range. The machine learning method assists in identifying the optimal structure of the AGNR structure and reveals that the C-13 atoms tend to form a continuous interface barrier perpendicular to the carrier transport direction to suppress the propagation of phonons through pure isotope-substituted AGNR structures. However, the *ZT* of DAGNR is 0.012 higher than that of the pristine C-12 structure. An important reason is that the C-12 is only one neutron less than C-13, the difference between isotope atoms is small. In theoretical simulations, it has been proven that the thermal conductivity of carbon nanomaterials can be further reduced by using pseudo atoms such as Carbon-22 [87].

Although optimization of periodical isotope-substituted AGNR structures performs a critical role in revealing the regulation mechanism of their thermal transport, accurate isotope atom site control is still challenging based on the available techniques. To obtain the optimal structure with larger *ZT*, we investigated the AGNR isotope superlattices composed of C-12 and C-13 and manipulated antidots at the interfaces (model B). The synthesis of isotope superlattices [36] and antidot structures [35] of graphene has been realized experimentally. In this part, 13-AGNR with *N*=3*p*+1 was selected to ensure the thermodynamic stability of the different antidot structures. The AGNR length remained at 25.56 nm with 14 design units in the central region. Figure 5a shows the period, optimal, isotope and antidot AGNR structures. The periodic AGNR structure indicated that the C-12 and C-13 units alternate in the central region and introduce antidots at the formed interfaces. The optimal structure is the candidate structure with the largest *ZT* processed by all machine learning algorithms. Separated isotope and antidot structures extracted from the optimal structure were also built up to exploit the contribution of isotopes and antidots in improving thermoelectric performance. The optimal structure is aperiodic with a *ZT* of 0.894, which is 5.69 times larger than that of the pristine structure (C-12). To verify the reliability of the tight-binding (TB) model taken in this work, first-principles calculations based on density generalized theory (DFT) were additionally adopted to analyze the electron transport properties of the pristine, periodic and optimal AGNRs structures. The generalized gradient approximation (GGA) with the Perdew-Burke-Ernzerhof functional (PBE) and cut-off energy of 150 Ry was used [88], and the obtained *ZT* values are shown in Fig. S6. The result shows that the *ZT* values of the three structures calculated based on the TB model and DFT have the similar trend of variation. Therefore, it is reasonable to choose TB to search the optimal structure. In Fig.5b, we compared the different properties of $Re$, $\sigma$, $S$, $Zp$ and $ZT$ of the pristine, period, optimal, isotope and antidot AGNR structures (five representative structures) in detail. The thermal resistance $Re$ is the reciprocal of the total thermal conductivity $k$, i.e. $Re = 1/k$. All the values were normalized by the pristine structure property. The thermal resistance of the periodic, optimal and antidot structures are all more than 2.50 times that of the pristine one, while the isotope structure is only 1.18 times, which demonstrates the superiority of AGNR with antidots in reducing the thermal conductivity. For the thermoelectric power factor, the optimal/antidot structure increases to 1.79 times, and retains 82.2% of the pristine electrical conductivity. However, the electrical conductivity of periodic structure is only about half that of the pristine one, as a result the increase of its thermoelectric power factor is rather limited. The $ZT$ of the period structure increases to 3.42 times, which is 2.47 times less than the optimal structure. Meanwhile, the $ZT$ of the antidot structure is close to 5.00 times, whereas the isotopic structure is only 1.18 times. Based on the above discussions, two major conclusions can be drawn: i) The further improvement of AGNR thermoelectric properties by manipulating the isotope atomic arrangements combined with antidots has shown great advantages of machine learning in designing structures with large degrees of freedom. ii) The antidot structure significantly improves AGNR thermoelectric properties by reducing thermal conductivity and increasing thermoelectric power factor compared to the isotope structure.

To further gain insights into the improvement mechanism of the thermoelectric properties of AGNRs, we investigated the phonon/electron transport properties. Figure 5c shows the phonon transmission functions of five representative structures. The AGNR structures with antidots obviously suppress the phonon transmission function. Unlike purely isotope effects, such phenomena occur almost over the entire phonon energy range. One reason is the reduction of the effective width of the AGNR conducting plane, due to the appearance of the antidot. Another major reason is the phonons scattering introduced by nanopores [31]. Moreover, the effective width of the electron transport in AGNRs is reduced by the antidots. Hence, the electron transmission functions of the periodic and optimal/antidot structures decrease compared to the pristine/isotope structure, but are not as significant as the phonon transmission functions, as shown in Fig. 5d. Remarkably, the five structures have a nearly identical values when the electron transmission function shifts from zero to the first step in the negative chemical potential region (marked with red circles in Fig. 5d). Combined with the observation of Fig. 5e, it reveals that the peak *ZT* of different structures is taken from this transition phase, which is favorable for maintaining the electrical conductivity. At the same time, the introduction of the antidots opens the band gap to obtain a large Seebeck coefficient [31]. The increase in *ZT* of the optimal structure is a joint effect of the electron and phonon transmission properties, of which the phonon transport suppression effect is relatively more effective due to the higher increase in thermal resistance.

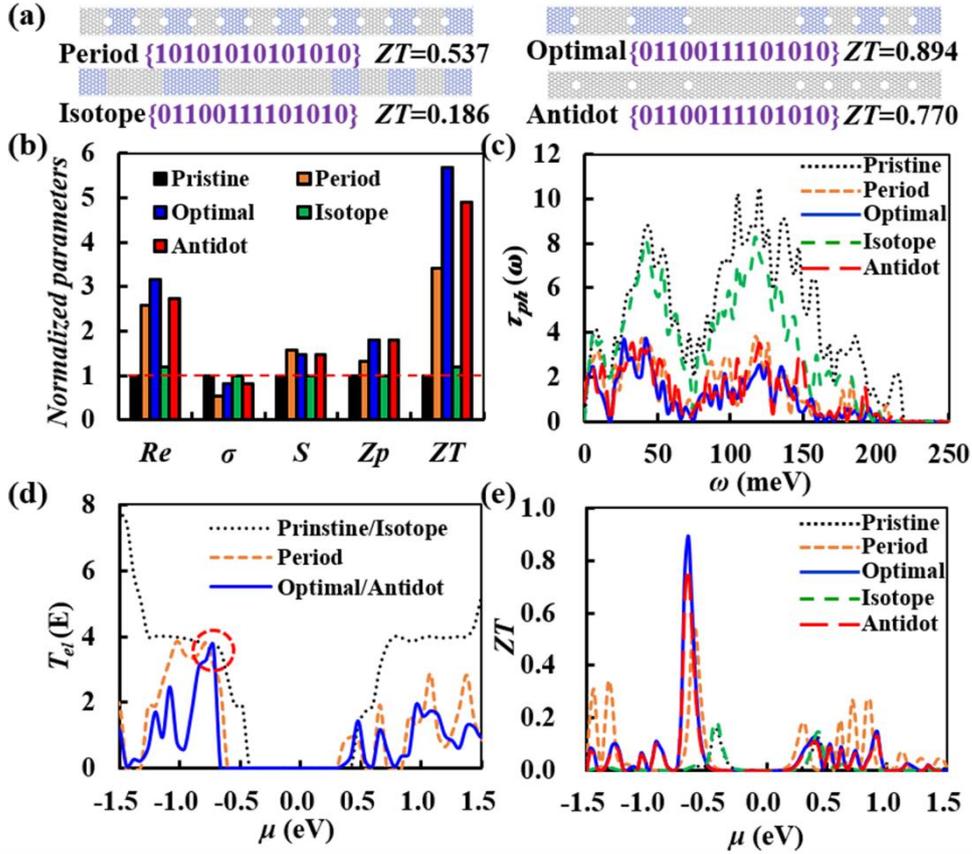

**Fig 5.** Optimization of AGNR superlattices with isotopes and antidots. (a) Period, optimal, isotope and antidot AGNR structures. (b) Phonon, electron and thermoelectric properties—$R_e$, $\sigma$, $S$, $Z_p$ and $ZT$—of five representative structures: pristine period, optimal, isotope and antidot AGNR structures. The values are normalized by the pristine structure. (c) and (d) Phonon/electron transmission functions of five representative structures. (e) Thermoelectric figure of merit (*ZT*) vs. chemical potential at *T* = 300 K of five representative structures. The isotope AGNR structure has the same isotope distribution as the optimal AGNR structure, but without the antidot. The antidot AGNR structure is exactly the opposite.

In this study, we also compared the first 15 AGNR structures in descending order of *ZT* ($ZT \geq 95\%Z_{max}T$), as shown in Fig. S7. From the inset on the bottom right, it shows that these structures have a moderate number of antidots, constant at 7 or 8. For the concentration of C-13, although the first four AGNR structures remain constant at 6/14, the later structures show relatively large fluctuation, between 4/14 and 8/14. This reflects the importance of antidots in model B for the optimization of the thermoelectric properties of AGNR structures. It has to be mentioned that the exact arrangement of carbon atoms and the elaborate design of antidot locations are essential. For instance, the candidate structure with descriptor {10010011101101} has the same number of antidots and C-13 as the optimal structure (top-1), but the *ZT* of 0.237 is about 1/4 of the optimal structure. Currently, the optimal structure has a *ZT* of 0.894, which is sufficient for low-power sensors applications [89], but not enough to compete with large scale energy systems in efficiency (*ZT* > 3) [90]. Previous studies have optimized the thermoelectric properties of similar scales of monolayer GNRs by introducing defects, creating interfaces or designing the topology, the *ZT* values are typically in the range of 0.1 to 0.8 at 300 *K* [38, 39, 91-94]. Also, the *ZT* values are influenced by parameters such as the length of the transport direction and the temperature [25]. Subsequently, we will increase the *ZT* of AGNRs by introducing heterojunction structures [88, 95]. The size of the optimal AGNR structure is 25.56 nm × 1.48 nm, and the nanopore diameter is 0.57 nm, whilst the state-of-the-art helium ion microscope has a super small beam spot as small as 0.5 nm [96]. Thus, the aperiodic isotopic superlattice preparation may be achieved via chemical vapor deposition technology by controlling pulsed signals to allow different isotopic carbon source gases into the reaction chamber. Then, the antidots can be prepared and sculpted to ideal nanoribbon sizes by helium ion beam etching. Of course, the precise combination of these technologies may still be challenging at present, but with the rapid development of advanced microfabrication and characterization techniques, it is believed that this will be possible in the near future.

## 4. Conclusions

To summarize, the NEGF method combined with machine learning algorithms has been demonstrated to accelerate the optimization of the thermoelectric properties of AGNRs. Compare to other optimization algorithms, BO shows high optimization efficiency, although it is comparable to GA in terms of optimization ability. In the optimization of pure isotope-substituted periodic AGNRs, we found the significant influence of high-frequency phonon modes in suppressing the thermal transport. However, the improvement of thermoelectric performance is limited due to the small differences between isotopic carbon atoms. To obtain the structure of AGNR with large *ZT*, we further inserted antidots at the interfaces of the isotopic AGNR superlattice structures, which could significantly improve the thermoelectric properties of AGNRs up to 5.69 times larger than that of the pristine structure. The obtained optimal superlattice structure is aperiodic, indicating that a detailed arrangement of isotopic carbon atoms combined with antidots is necessary. The comparison of isotope and antidot structures shows that the antidot structure significantly improved AGNR thermoelectric performance by reducing thermal conductivity and increasing thermoelectric power factor. Overall, the proposed machine learning guided manipulation of disordered antidots and interfaces has shown great advantages in enhancing the thermoelectric properties of low-dimensional nanostructures with large degrees of freedom.

**Declaration of competing interest**

The authors declare that they have no known competing financial interests or personal relationships that could have appeared to influence the work reported in this paper.

**Acknowledgements**

This work was supported by the National Natural Science Foundation of China (No. 52006134), Shanghai Pujiang Program (No. 20PJ1407500), Shanghai Key Fundamental Research Grant (No. 21JC1403300), and the National Key R&D Program of China (2021YFB3702300). The computations in this paper were run on the $\pi$ 2.0 cluster supported by the Center for High Performance Computing at Shanghai Jiao Tong University. This research was also partially supported by Open Research Fund of CNMGE Platform & NSCC-TJ.

Supplemental Materials for

# Enhancing Thermoelectric Properties of Isotope Graphene Nanoribbons via Machine Learning Guided Manipulation of Disordered Antidots and Interfaces

**Xiang Huang[1], Shengluo Ma[1], Haidong Wang[2], Shangchao Lin[1], C. Y. Zhao[1], Hong Wang[3], and Shenghong Ju[1, 3, *]**

[1] China-UK Low Carbon College, Shanghai Jiao Tong University, Shanghai, 201306, China

[2] Key Laboratory of Thermal Science and Power Engineering of Ministry of Education, Tsinghua University, Beijing, 100084, China

[3] Materials Genome Initiative Center, School of Material Science and Engineering, Shanghai Jiao Tong University, Shanghai 200240, China

---

[*] Corresponding author: shenghong.ju@sjtu.edu.cn (S.H. Ju)

## Section S1. Nonequilibrium Green's function (NEGF) method for electron/phonon transport properties analysis

We utilized the Nonequilibrium Green's function (NEGF) method to obtain the electron/phonon transmission functions of AGNRs [1-6].

For electron transport analysis, the retarded Green's function in the central region is calculated as:

$$G^r(E) = \left[EI - H_C - \sum\nolimits_L^r(E) - \sum\nolimits_R^r(E)\right]^{-1}, \quad (S1)$$

where $E$ is the energy of the incident electron, $I$ is the identity matrix, $H_C$ is the Hamiltonian matrix, $\sum_{L,R}(E)$ is the self-energy of the left/right lead. The electron transmission function of the central region is:

$$\mathcal{T}(E) = Tr[\Gamma_L(E)G^r(E)\Gamma_R(E)G^a(E)], \quad (S2)$$

where the level broadening matrices $\Gamma_L(E) = i(\sum_L^r(E) - \sum_L^a(E))$ and $\Gamma_R(E) = i(\sum_R^r(E) - \sum_R^a(E))$ describe the rates of inflow from the left lead and outflow into the right lead. The electronic current $I_c$ is given by the Landauer formula:

$$I_c = \frac{2e}{h}\int_{-\infty}^{\infty} dE\mathcal{T}(E)[f(E,\mu_L) - f(E,\mu_R)], \quad (S3)$$

where $f(E,\mu_{L,R}) = \left(e^{E-\mu_{L,R}/k_BT} + 1\right)^{-1}$ is the Fermi-Dirac distribution function on the left and right leads with chemical potentials $\mu_{L,R}$. The voltage difference between left and right is $V = (\mu_L - \mu_R)/e$. The linear electronic conductance $\sigma(\mu) = dI_c/dV$ is written as

$$\sigma(\mu) = \frac{2e^2}{h}\int_{-\infty}^{\infty} dE\mathcal{T}(E)\left(-\frac{\partial f(E,\mu)}{\partial E}\right) = e^2 L_0, \quad (S4)$$

For phonon transport analysis, it can be calculated mathematically similarly to electron transport, using the substitutions:

$$EI \to \omega^2 M, \quad H_C \to K_C. \quad (S5)$$

The transmission function through the central region is given by:

$$\tau(\omega) = Tr[\Gamma_L(\omega)G^r(\omega)\Gamma_R(\omega)G^a(\omega)], \quad (S6)$$

where ω is the phonon energy, $M$ is the dynamic matrix composed of element masses, and $K_C$ is the dynamic matrix of the central region. The final thermal conductance is given by,

$$\kappa_p(T) = \frac{J_{ph}}{\Delta T} = \frac{\hbar^2}{2\pi k_B T^2} \int_0^\infty \omega \tau(\omega) \frac{e^{\frac{\hbar\omega}{k_B T}}}{\left(e^{\frac{\hbar\omega}{k_B T}} - 1\right)^2} d\omega, \tag{S7}$$

where $k_B$ is the Boltzmann constant and $\hbar$ is the reduced Plank constant.

For thermoelectric properties, the dimensionless figure of merit $ZT$ is calculated by:

$$ZT = \frac{S^2 G_e T}{\kappa_p + \kappa_e}, \tag{S8}$$

where $\sigma$ is the electrical conductivity, $S$ is the Seebeck coefficient, $T$ is the absolute temperature, $k_p$ and $k_e$ are the lattice and electronic thermal conductivity, respectively. Moreover, the Seebeck coefficient and electronic thermal conductivity can be calculated as follows:

$$S(\mu, T) = \frac{1}{eT} \frac{L_1(\mu)}{L_0(\mu)}, \tag{S9}$$

$$\kappa_e(\mu) = \frac{1}{T}\left\{L_2(\mu) - \frac{[L_1(\mu)]^2}{L_0(\mu)}\right\}. \tag{S10}$$

We can get the function $L_m(\mu)$ from electron transmission function,

$$L_m(\mu) = \frac{2}{h} \int_{-\infty}^\infty dE \mathcal{T}(E)(E-\mu)^m \left(-\frac{\partial f(E,\mu)}{\partial E}\right). \tag{S11}$$

**Section S2. Width effect on thermoelectric properties of AGNRs**

Figure S1 shows the phonon transport properties of AGNRs with different widths (7-AGNR, 8-AGNR and 9-AGNR, Length was 0.426 nm), which indicates the dependence of the phonon thermal conductivity on the width in the system. As the width decreases, the phonon transmission function is usually suppressed. In the case of narrower AGNRs, the edge-localized phonon effect plays a leading role, which is weakened with the increases in AGNR width and leads to an increase in thermal conductivity [7]. Moreover, the lattice thermal conductivity of AGNRs increase with the increasing temperature. This is because more high-frequency phonon modes are excited, which is beneficial for thermal transport [8].

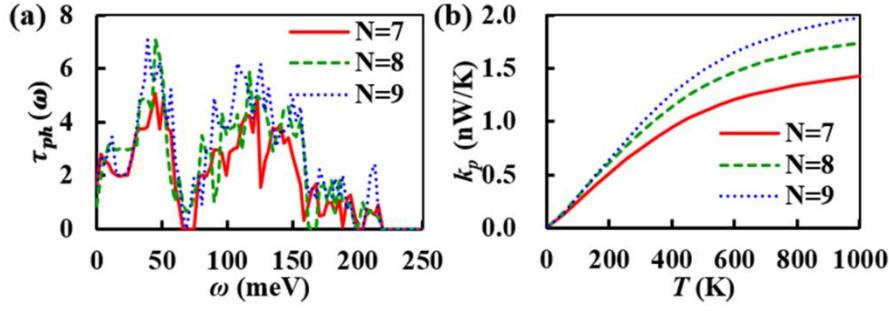

**Fig. S1.** Phonon transport properties of AGNRs with different widths (Length was 25.56 nm). (a) Phonon transmission function. (b) Lattice thermal conductivity.

Fig.S2 shows the electron transport properties of AGNR. The chemical potential at which the density of states is zero corresponds to the band gap of the device. The 8-AGNR with $N=3p+2$ exhibits metallic properties, which has the smallest energy bandgap and peak of Seebeck coefficient [9]. While 7-AGNR and 9-AGNR behave as semiconductors, the energy bandgap and peak of Seebeck coefficient are larger than that of metallic graphene respectively and are inversely proportional to the width [10, 11]. The electron conductivity reflects a consistent character with the electron transmission function, and its rise is caused by small peaks in the device density of states.

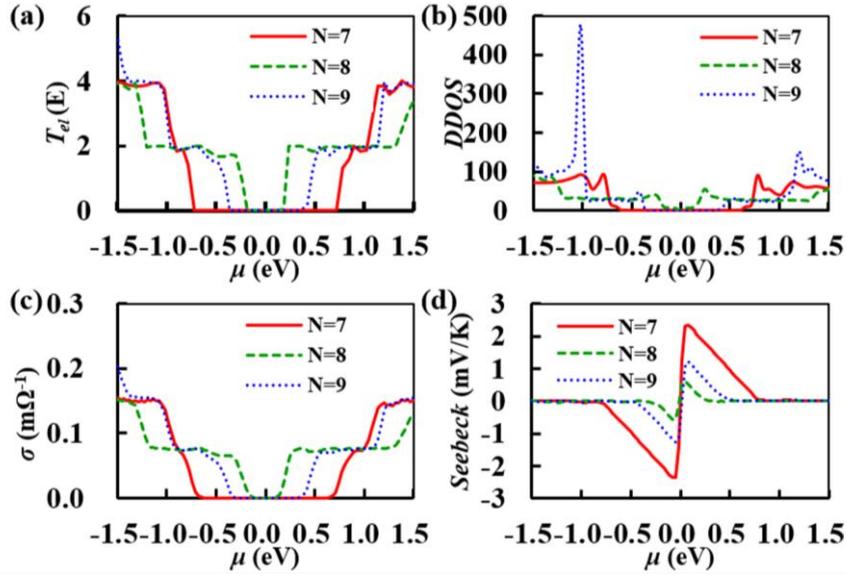

**Fig. S2.** Electron transport properties of AGNRs with different widths (Length of 25.56 nm). (a) Electron transmission function. (b) Device density of states. (c) Electron conductivity. (d) Seebeck coefficient.

The thermoelectric properties of AGNRs are affected by the coupling of electrical conductivity $\sigma$, Seebeck coefficient $S$ and thermal conductivity $k$, as shown in Fig. S3. It is worth mentioning that the power factor $Z_p$ peak positions correspond exactly to the electron conductance transitions from one step to another, rather than the positions of the peak of *S* [6]. Therefore, 8-AGNR has a larger $Z_p$ peak although it has a smaller peak of *S*. The *ZT* peak of 7-AGNR, 8-AGNR and 9-AGNR are 0.245, 0.242 and 0.147, respectively. The 7-AGNR has a larger *ZT* due to its low thermal conductivity. In addition, ignoring lattice thermal conductivity, the $ZT_e$ peak of 7-AGNR and 8-AGNR are about 706.264 and 21.204, which are 2882.6 times and 87.6 times of the corresponding *ZT*, respectively. The difference between $ZT_e$ and *ZT* reflects the significant contribution of lattice thermal conductivity to the total thermal conductivity. Furthermore, the AGNR with *N*=3*p*+1 is more sensitive in the same sequence (the integer *p* is equal). Based on the above, we chose the AGNRs with *N*=3*p*+1 for the follow-up research.

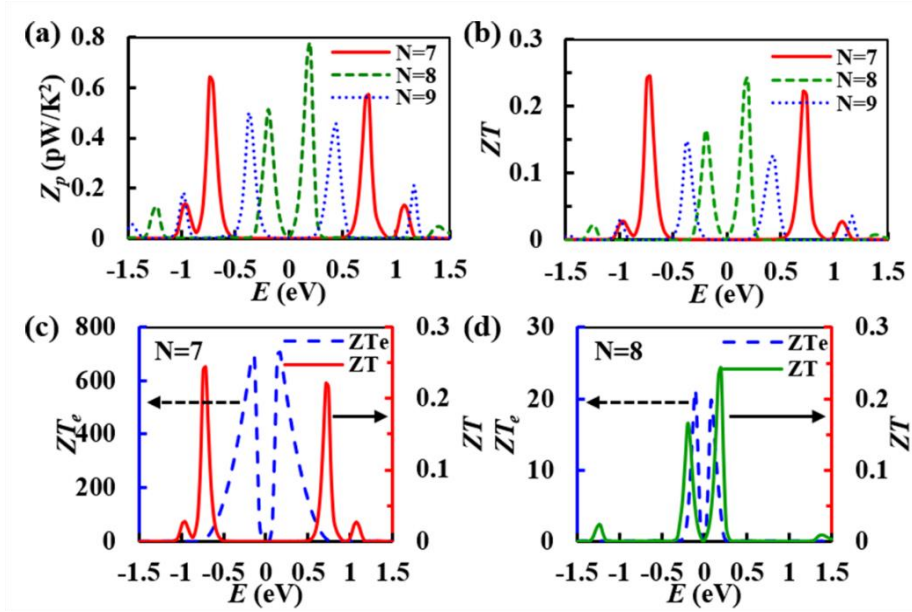

**Fig. S3.** Thermoelectric properties of AGNRs with different widths (Length is about 25.56 nm). (a) Thermoelectric power factor. (b) Thermoelectric figure of merit. (c) and (d) Effect of lattice thermal conductivity on the thermoelectric figure of merit for 7-AGNR and 8-AGNR.

**Section S3. Comparison of machine learning algorithms**

Apart from the Bayesian optimization (BO) and genetic algorithm (GA) algorithms, we additionally performed Monte Carlo tree search (MCTS) [12] and random optimization (RO) algorithms for comparison. MCTS is a random best-first tree search algorithm, which iteratively searches the candidate space systematically to obtain the maximum of the black-box function [13], and the process includes four steps: 1) Selection, the tree is traversed from the root to the most promising leaf using a comparative score; 2) Expansion, the selected node is expanded by adding a child node; 3) Simulation, the complete descriptors of candidate structure is generated randomly, and calculate its ZT; 4) Back-propagation, the information of node is updated along the path back to the root. The upper confidence bound applied to trees (UCT) was used to evaluate child nodes when traversing the tree [14],

$$UCT = \frac{z_k}{v_k} + C\sqrt{\frac{2\ln v_p}{v_k}}, \tag{S12}$$

where $v_k$ is the traversal times of the node, $v_p$ is the traversal times of the parent node, $z_k$ is the accumulated merit of the node and *C* is constant to balance exploration and exploitation. The constant C can be adjusted according to experience, and the value corresponds to the search breadth and depth.

As depicted in Fig. 5, 50 rounds of optimization were performed with different initial choices of AGNR structures, and 2000 structures were calculated in each round. We performed the optimization of periodic AGNR structures within single and double units in model A, respectively, named model A-S and model A-D. From the convergence of the curves, the BO exhibits a superior optimization ability in model A. However, in model B, the BO and GA have comparable optimization ability, which is better than MCTS and RO. To enable a better evaluation of the different algorithms, we defined the optimization ability and efficiency, based on the statistics of the first five AGNR structures corresponding to the ZT value, respectively. Details are discussed in the main text.

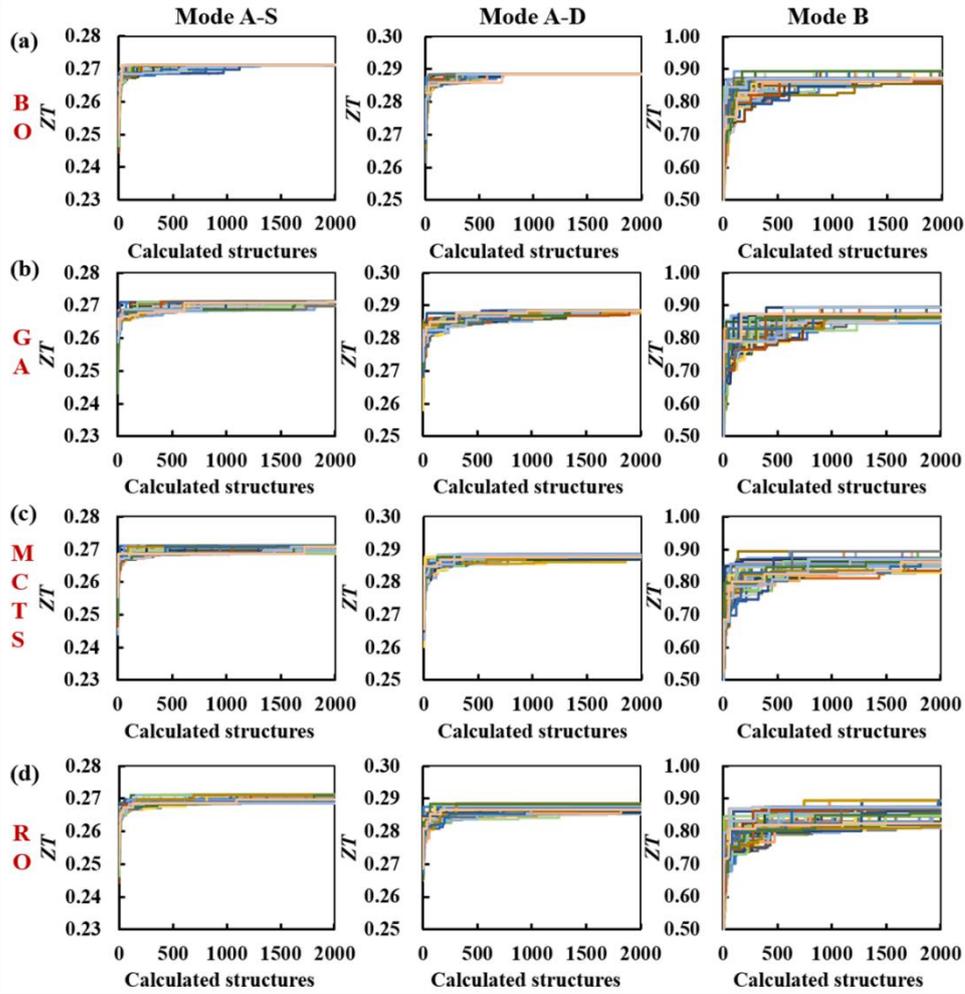

**Fig. S4.** The 50 rounds of optimization with different initial choices of AGNR structures. (a) Bayesian optimization (BO). (b) Genetic algorithm (GA). (c) Monte Carlo tree search (MCTS). (d) Random optimization (RO).

**Section S4. Effect of isotopic carbon atoms distribution on thermoelectric properties of AGNR structures**

To investigate the effect of isotopic carbon atoms distribution on thermal conductivity/thermoelectric properties, three typical structures in Fig. S5a were selected for analysis, where the {00110111011001} (Type-I) and {01100111000000} (Type-II) are the configurations of AGNRs with minimum and maximum $ZT$ from the system of single unit regulating DOF, while the {11000000000111} (Type-III) is the configuration with minimum $ZT$ with the same C-13 concentration as Type-II (C-13 concentration of 9/14). For the phonon transport calculations, the electrodes and extension regions are uniformly set to C-12. The Type-I AGNR has the largest thermal conductivity, which intuitively provides a continuum path of C-12 for phonons to coherently propagate. Similarly, although the Type-III AGNR fixes the C-13 concentration, it still maintains a continuous path of C-12 for coherent phonon propagation and has the maximum thermal conductivity at this concentration (C-13 concentration of 9/14). Type-II AGNR is the optimal structure with the largest *ZT* and the smallest thermal conductivity, whose C-13 atoms form a continuum interface barrier in the direction perpendicular to the transport, suppressing the propagation of phonons. The above results are consistent with previous studies on the thermal conductivity of Si/Ge interface structures [15]. In addition, the phonon transmission in Fig. S5b reflects the suppression of phonon propagation by the distribution of the C-13 atoms in type-II AGNR.

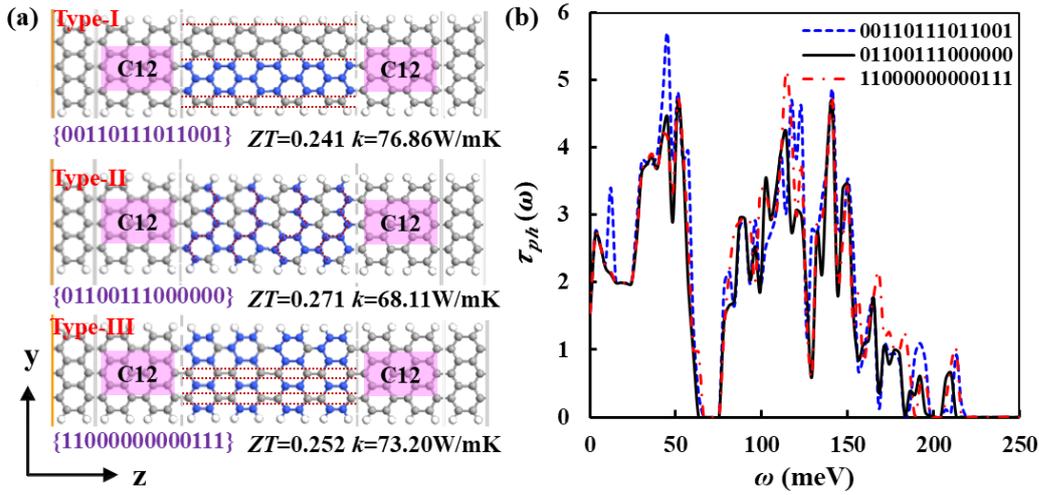

**Fig. S5.** Effect of isotopic carbon atoms distribution on thermal conductivity/ thermoelectric properties of AGNR structures. (a) and (b) Configurations and phonon transmission of three typical structures. {00110111011001} and {01100111000000} are the configurations of AGNRs with minimum and maximum ZT from the system of single unit regulating DOF. {11000000000111} is the configuration with minimum ZT at a fixed C13 concentration of 9/14.

**Section S5. Reliability verification of TB model-based calculation results**

First-principles calculations based on density generalized theory (DFT) were additionally adopted in Quantum ATK software to analyze the electron transport properties of the pristine, periodic and optimal structures of model B. The generalized gradient approximation (GGA) with the Perdew-Burke-Ernzerhof functional (PBE) and cut-off energy of 150 Ry was used [16], the obtained *ZT* values are shown in Fig. S6. The result shows that the *ZT* values of the three structures calculated based on the TB model and DFT have the similar trend of variation. Therefore, it is reasonable to choose TB to search the optimal structure.

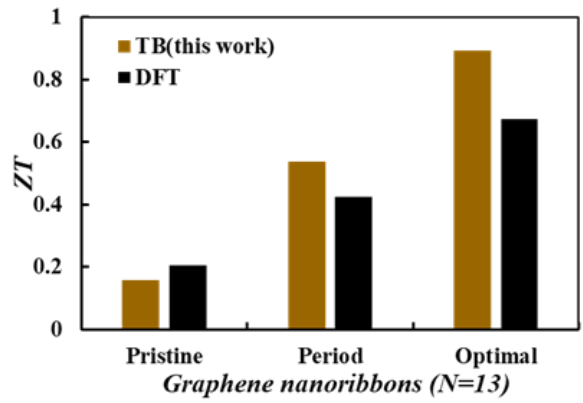

**Fig. S6.** Comparison of *ZT* values based on TB model and DFT calculation.

**Section S6. The first 15 AGNR structures in descending order of *ZT*, with $ZT \geq 95\% Z_{max}T$**

Figure S7 shows the first 15 AGNR structures in descending order of *ZT*, with $ZT \geq 95\% Z_{max}T$. Form the inset on the bottom right, we can find that these structures have a moderate number of antidots, constant at 7 or 8. For the concentration of C-13, although the first four AGNR structures remain constant at 6/14, the later structures have relatively large fluctuation, between 4/14 and 8/14. It reflects the importance of antidots in model B for the optimization of the thermoelectric properties of AGNR structures. Of course, the exact arrangement of carbon atoms and the elaborate design of antidot locations are essential. For instance, the candidate structure with descriptor {10010011101101} has the same number of antidots and C-13 as the optimal structure (top-1), but the *ZT* of 0.237 is about 1/4 of the optimal structure.

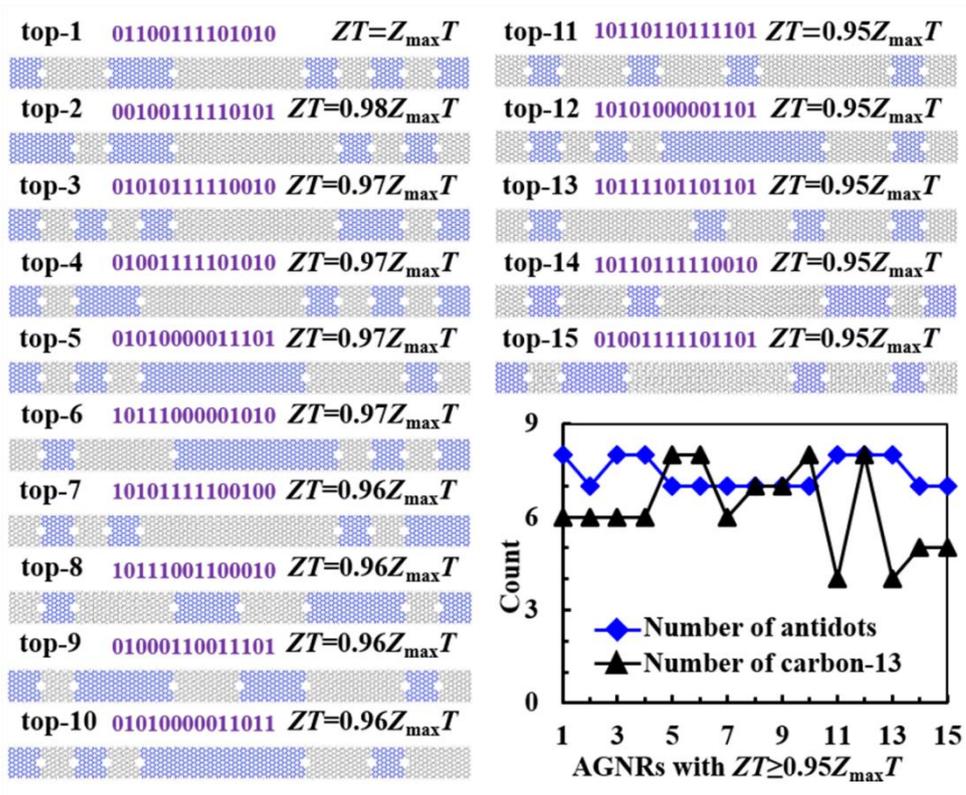

**Fig. S7.** First 15 AGNR structures in descending order of *ZT*, with $ZT \geq 95\% Z_{max}T$. The inset on the bottom right illustrates the statistics of the number of antidots and carbon 13 for different structures, whose horizontal coordinates correspond to the descending order of *ZT*.